\documentclass[lettersize,journal]{IEEEtran}
\IEEEoverridecommandlockouts
\usepackage{amsmath,amsfonts}
\usepackage{algorithmic}
\usepackage{algorithm}
\usepackage{array}
\usepackage{subfigure}
\usepackage{textcomp}
\usepackage{stfloats}
\usepackage{url}
\usepackage{subcaption}
\usepackage{verbatim}
\usepackage{graphicx}
\usepackage{cite}
\usepackage{bm}
\usepackage[usenames]{color}
\usepackage{xspace,colortbl}
\begin{document}
\bstctlcite{setting}

\title{Digital Twin Online Channel Modeling: Challenges, Principles, and Applications}

\author{Junling Li,~\IEEEmembership{Member,~IEEE}, Cheng-Xiang Wang,~\IEEEmembership{Fellow,~IEEE}, Chen Huang,~\IEEEmembership{Member,~IEEE}, \\ Tianrun Qi,~\IEEEmembership{Student Member,~IEEE}, and Tong Wu

\thanks{This work was supported by the National Natural Science Foundation of China (under Grants 62394290, 62394291, and 62301151) and the Start-up Research Fund of Southeast University under Grant RF1028623062. (Corresponding authors: Cheng-Xiang Wang and Chen Huang.)}
\IEEEcompsocitemizethanks{ \IEEEcompsocthanksitem Junling Li and Cheng-Xiang Wang are with the National Mobile Communications Research Laboratory, School of Information Science and Engineering, Southeast University, Nanjing 211189, China and also with the Pervasive Communication Research Center, Purple Mountain Laboratories, Nanjing 211111, China (email: junlingli, chxwang@seu.edu.cn).
\IEEEcompsocthanksitem Chen Huang is with the Pervasive Communication Research Center, Purple Mountain Laboratories, Nanjing 211111, China, and also with the National Mobile Communications Research Laboratory, School of Information Science and Engineering, Southeast University, Nanjing 211189, China (e-mail: huangchen@pmlabs.com.cn). 
\IEEEcompsocthanksitem Tianrun Qi and Tong Wu are with National Mobile Communications Research Laboratory, School of Information Science and Engineering, Southeast University, Nanjing 210096, China (email: tr\_qi, wutong@seu.edu.cn).}}

\maketitle{}

\begin{abstract}
Different from traditional offline channel modeling, digital twin online channel modeling can sense and accurately characterize dynamic wireless channels in real time, and can therefore greatly assist 6G network optimization. This article proposes a novel promising framework and a step-by-step design procedure of digital twin online channel models (DTOCM). By enabling continuous visualization and accurate prediction of dynamic channel variations, DTOCM can synchronize the performance between simulated and real networks. We first explore the evolution and conceptual advancements of DTOCM, highlighting its visions and associated challenges. Then, we explain its operational principles, construction mechanisms, and applications to typical 6G scenarios. Subsequently, the real-time channel information provisioning and visualization capabilities of DTOCM are illustrated through our DTOCM platform based on practical scenarios. Finally, future research directions and open issues are discussed.

\end{abstract}


\section{Introduction}

\IEEEPARstart{S}{ixth} generation (6G) wireless communication networks are expected to have “global coverage, all spectra, full applications, all senses, all digital, and strong security" \cite{6G_CXW}. Traditional network optimization often relies on costly and time-consuming channel measurements, trial-and-error processes, and engineering experience. Consequently, most networks can only achieve about 60$\%$ of their potential performance, leaving significant room for improvement \cite{ZQluo}. Network simulation that can accurately characterize the real network performance is crucial for guiding the deployment of 6G networks and enhancing their performance. However, simulated network performance in the lab often does not match real network performance. The reasons are mainly two-fold: First, the channel models and parameters used in the lab simulation do not match the real dynamic propagation environment, since channel measurements have not been conducted in the tested scenario in the real network. Second, even if a channel model that well fits the tested scenario exists, its parameters still cannot adapt to time-varying tested scenarios, resulting in a mismatch between the synthetic channel in the lab simulation and the practical wireless channel in the real network \cite{Wang20channel}. 

Digital twin (DT) technology refers to the creation of a virtual replica of a physical system or environment that is updated continuously with real-time data \cite{Tao2019}. In communication networks, digital twins mirror real-world network conditions, enabling simulation, analysis, and optimization of performance without affecting actual operations. Recently, a few studies leveraging the DT technology to address the issue of mismatch between simulated network performance and real network performance have been reported \cite{wang2023}\cite{zhang2023}. In \cite{wang2023}, an environmentally aware DT platform has been presented to achieve channel prediction based on environmental semantics. However, the study lacks a detailed discussion on channel modeling methods and online updating methods, overlooking the crucial role of channel modeling in DT online channel platforms. In \cite{zhang2023}, twin data with channel statistical characteristics has been generated using generative adversarial networks, achieving a close match with real-world channels. However, this method has limited applicability and cannot be easily extended to digital twin online channel models in different scenarios. Therefore, there exists an urgent need for digital twin online channel models (DTOCM) that can reflect the channel characteristics of the real network communication environment in real time. 

DTOCM allows for continuous monitoring and real-time analysis of channel conditions, providing visualizations and predictions of dynamic channel variations \cite{Alkhateeb}. As 6G networks move towards supporting highly dynamic environments, such as vehicular and unmanned aerial vehicle (UAV)-aided communication systems, DTOCM can accurately characterize dynamic channels in these environments, providing detailed and real-time insights into how signals propagate in the presence of moving vehicles, changing road conditions, and varying traffic densities. Moreover, DTOCM enables real-time adaptation of channel parameters to optimize performance, which is crucial for maintaining reliable and efficient vehicular communication networks \cite{Lin2023}.



The feasibility of DTOCM is underpinned by two key factors. First, the close coupling relationship between the physical environment and channel characteristics can be partially captured through physical positioning and environmental features. This inherent correlation indicates that similar physical positions and environmental characteristics often yield similar channel characteristics, making the extensive data generated by base stations reusable. This reuse potential significantly aids in creating accurate digital twins for channel modeling. Second, advancements in positioning and perception accuracy, driven by improved communication networks and expanded frequency bands, enhance the ability to precisely locate and understand the environment. These enhanced capabilities enable more detailed and accurate digital representations of the wireless channels, aligning perfectly with the objectives of DTOCM. Together, these factors create a conducive environment for constructing DTOCM, offering four key benefits: 
\begin{itemize}
    \item \textbf{Reduced pilot overhead}: DTOCM enables the preloading of basic channel state information (CSI), allowing devices to obtain channel statistics while moving through a continuous 3D space. This can be achieved by accessing a database or through calculations based on position, orientation, and antenna parameters. By addressing communication network design challenges directly within channel research, DTOCM significantly reduces pilot overhead, enhancing network performance in the 6G era;
    \item \textbf{Real-time CSI provisioning and prediction}: DTOCM excels in real-time channel information provisioning and prediction, adapting to dynamic network conditions. This capability ensures that the 6G communication networks remain responsive to environmental changes and helps calibrate the performance between channel model simulations and actual environmental conditions, leading to more reliable and robust network performance;
    \item \textbf{Visualized channel information}: DTOCM provides visualization of channel information variations, predicting and highlighting changes in the communication environment. It allows for the display of node positions and motion states and offers real-time insights into channel characteristics and other relevant information. This visibility enhances environmental awareness and decision-making within future 6G networks;
    \item \textbf{Optimized network performance}: DTOCM can simulate scenarios in a virtual environment, providing insights and feedback to optimize real-world communication network deployment. This enables comprehensive testing in a controlled virtual setting, allowing rapid iterations and refinements. By leveraging the insights gained from these simulations, future 6G networks can be optimized effectively, balancing performance and efficiency. 
\end{itemize}


Foreseeing such promising benefits, this article proposes a novel digital twin online channel modeling framework, enabling real-time perception of dynamic environments by utilizing physical environmental sensors with various perception methods. This establishes a mapping between actual communication environment parameters, channel parameters, and channel characteristics. In addition, by integrating environmental perception with channel modeling, DTOCM enhances the prediction accuracy of changes in the wireless propagation environment, enabling more precise simulation of real-world wireless propagation scenarios and providing robust support for communication network optimization and decision-making.

This article is organized as follows. In Section \ref{SecII}, we explore the evolution from offline channel map to DTOCM, along with its visions and challenges. Then, in Section \ref{Sec-framework}, we present the proposed digital twin online channel modeling framework in detail, including its overall working principle, three-step construction mechanism, and typical applications in 6G communication scenarios. In Section \ref{Sec-visual}, we illustrate the feature of real-time provisioning and visualization of channel information of the proposed digital twin online channel modeling framework. Finally, we discuss some open research issues related to DTOCM in Section \ref{open_issue}, followed by Section \ref{conclusion} that concludes this article.

\section{The Evolution From Offline Channel Maps to DTOCM}
\label{SecII}
Channel maps have gained significant attention in recent years due to their powerful capability of describing the transmission characteristics of wireless channels, supporting the output of statistical channel characteristics offline, and providing information on nodes' location and movement status in the network \cite{Zeng2024}.

\begin{figure}[t]
    \centering
    \includegraphics[width=3.0in]{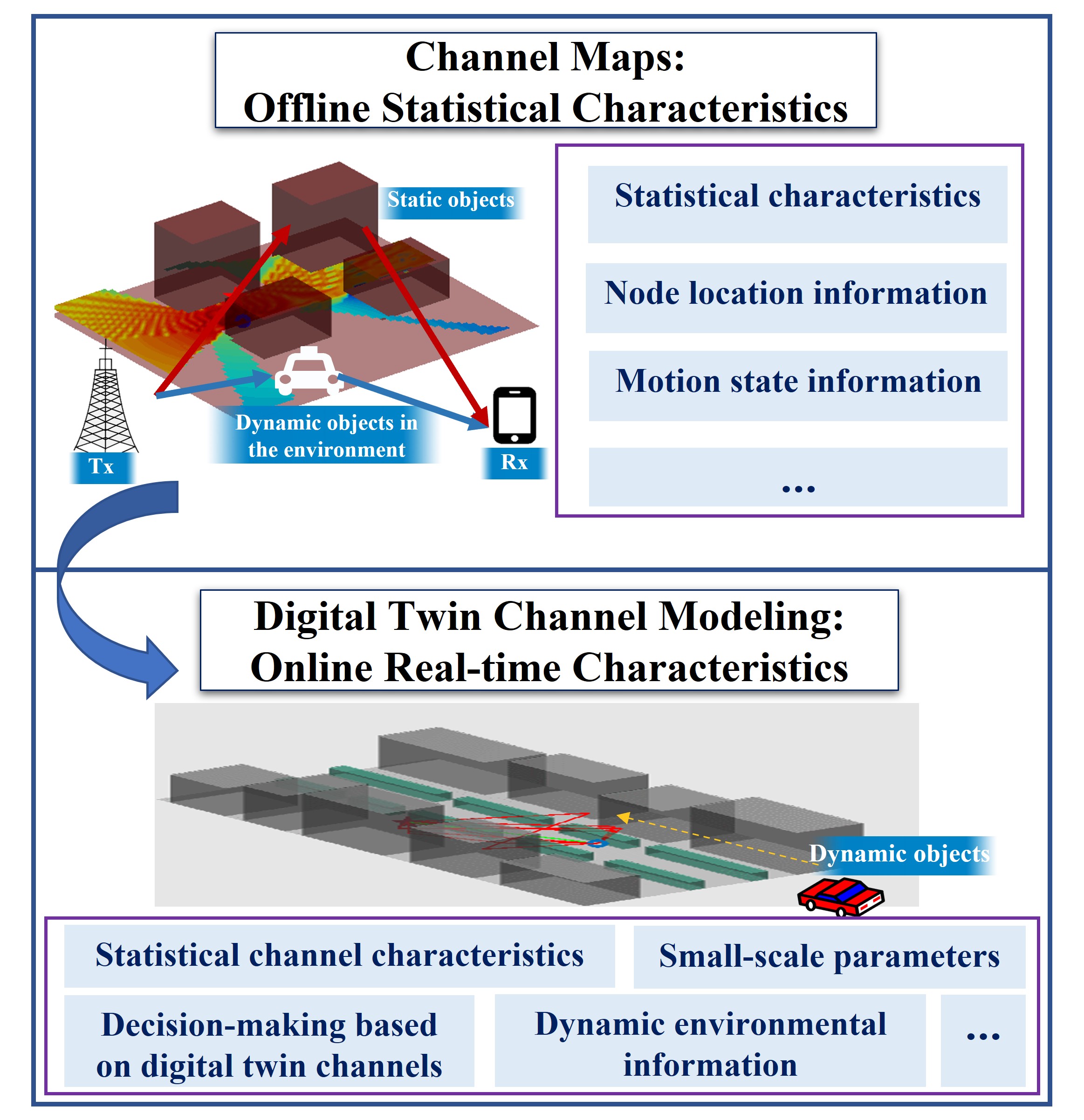}
    \caption{The evolution from offline channel maps to DTOCM.}
    \label{fromto}
\end{figure}

As shown in Fig. \ref{fromto}, being evolved from wireless channel maps, digital twin channels perform virtual mapping of physical systems, including communication networks, through artificial intelligence (AI) prediction and completion. This enables the digital twin modeling of wireless communication networks, which supports a better understanding and real-time provisioning and prediction of the wireless signal transmission process, thereby offering more accurate guidance for the design and optimization of 6G communication networks.

\subsection{Visions of DTOCM}
\begin{figure*}[!t]
	\centering 
        \includegraphics[width=\textwidth]{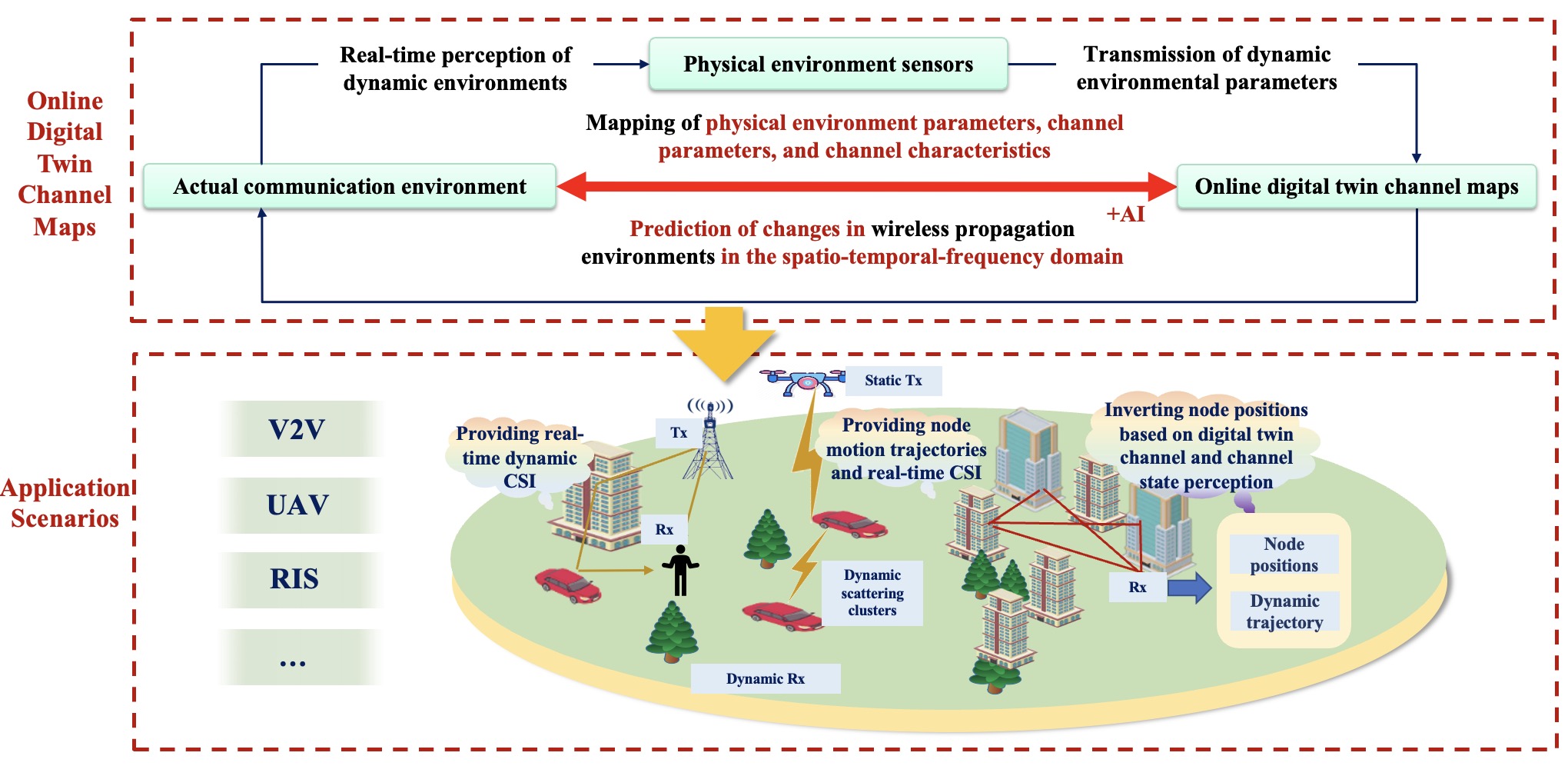} \\
	\caption{Visions and typical application scenarios of DTOCM.}
 \label{vision}
\end{figure*}

As illustrated in Fig. \ref{vision}, the DTOCM captures changes in CSI within real dynamic environments, displaying the positions and movement of objects and nodes while outputting customized channel parameters at both large and small scales. The DTOCM can facilitate real-time channel information prediction and assist in communication network optimization decisions. By leveraging physical environment sensors, the digital twin channel offers real-time awareness of dynamic environments, mapping actual communication environment parameters to channel parameters and characteristics. Combined with AI algorithms, it can predict changes in wireless propagation environments across spatial, temporal, and frequency domains.

The above visions can be elucidated in three typical application scenarios in 6G networks, as shown in Fig. \ref{vision}. In the first scenario, the transmitter and receiver are stationary while the scattering environment changes dynamically. The DTOCM can display real-time CSI such as amplitude, phase, delay, Doppler, and angle. In the second case, the transmitter and receiver are mobile in a dynamic environment. The DTOCM provides real-time CSI and the positions and movement of various nodes. In the third case, a mobile entity like a drone is in motion. The DTOCM can infer the drone's position based on changes in channel information.

\subsection{Challenges of DTOCM}
Exploring methods for obtaining DTOCM is still in its early stages, and there are still some technical challenges to address before achieving its visions: 
\begin{itemize}
    \item \textbf{Complex environment reconstruction}: The actual propagation environment is complicated, with high data collection costs, making it difficult to reconstruct dynamic twin environments. Therefore, it is imperative to explore efficient, low-cost, and high-precision methods for dynamically reconstructing twin environments. Multi-source data acquisition and integration could be a promising solution, where data from various sources (such as sensors, satellite imagery, LiDAR, and millimeter-wave radar) are first gathered and then integrated to create a comprehensive representation of the environment. Following that, AI algorithms can be applied to identify and categorize different environmental scenarios based on the data;
    \item \textbf{Non-ideal measurement condition}: Incomplete and non-ideal channel measurement conditions create data gaps in the channel modeling process and lead to inaccuracies in representing the actual channel conditions. Therefore, there is an urgent need to explore effective theories and algorithms to compensate for the lack of ideal measurement data. By implementing an AI-based spatiotemporal frequency prediction and completion method, DTOCM can effectively alleviate non-ideal measurement conditions. Based on the known measurement data at specific times, frequencies, and locations, ML methods such as LSTM, CNN, and MLP can be utilized to predict and fill in the unknown measurement data for unobserved times, frequencies, and locations, leading to more accurate and robust channel modeling and network optimization; 
    \item \textbf{AI database dependency}: Current AI-based wireless channel models rely heavily on databases, highlighting the urgent need for new AI-based predictive channel modeling methods. These new methods should offer improved generalization ability and broader applicability to enable more efficient and accurate channel characterization.
\end{itemize}

Overall, the current exploration of DTOCM construction methods is still in its early stages, and it is imperative to develop efficient approaches for intelligent channel characterization, addressing the accuracy issues in simulating real environments and predictive decision-making. Facing such challenges, we propose a promising digital twin online channel modeling framework detailed in following sections.

\section{Proposed Digital Twin Online Channel Modeling Framework}
\label{Sec-framework}
\subsection{Overall Principles}
\begin{figure*}[!t]
  \centering
  \includegraphics[width=0.9\textwidth]{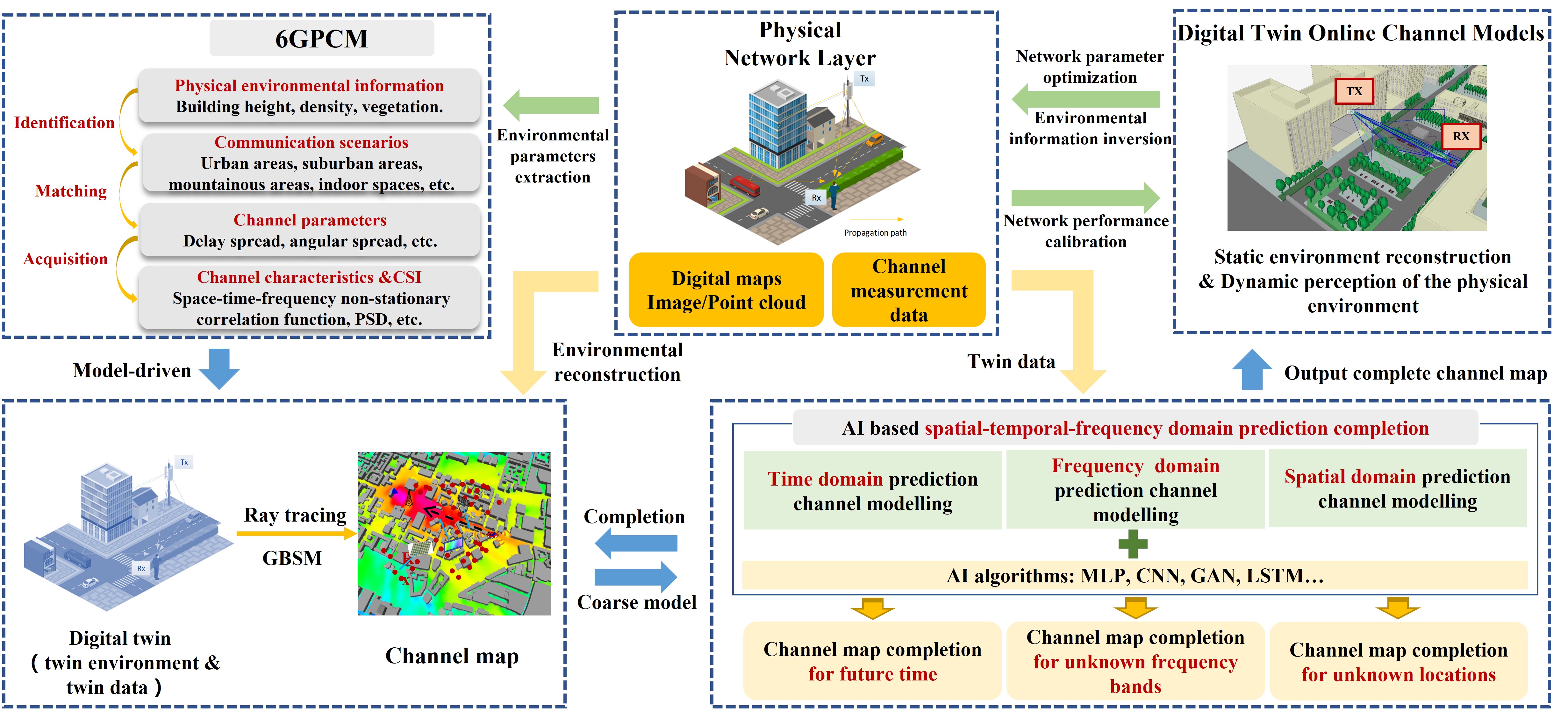} 
  \caption{Framework of the proposed digital twin online channel modeling.}
  \label{framework}
\end{figure*}

As shown in Fig. \ref{framework}, DTOCM involves comprehensive and exclusive classification of 6G scenarios, followed by real-time perception of specific communication scenarios through environmental information. This is then paired with the model parameters of the 6G pervasive channel model (6GPCM) \cite{6GPCM} to achieve data-driven communication scenario identification and model-driven full-scenario channel simulation.

Environmental and channel measurement data are first collected at the physical network layer for environmental reconstruction and data twinning, creating a digital twin layer. The digital twin layer utilizes ray tracing (RT) and geometric-based stochastic modeling (GBSM) to generate a wireless channel map. This map serves as a coarse model for AI prediction, where AI-based models leverage known-time channel databases to achieve effective and accurate spatio-temporal-frequency predictions \cite{huangchen}. These predictions fill gaps in the wireless channel map, such as unknown scenarios, future timeframes, and unfamiliar frequency bands. Through this process, a relatively complete digital twin channel map can be achieved. This map can be used for network parameter optimization and environmental information inversion at the physical network layer. Additionally, the physical network layer can provide the digital twin channel map with the necessary parameters or data for network performance calibration.


\begin{figure*}[t]
    \centering
    \includegraphics[width=0.95\textwidth]{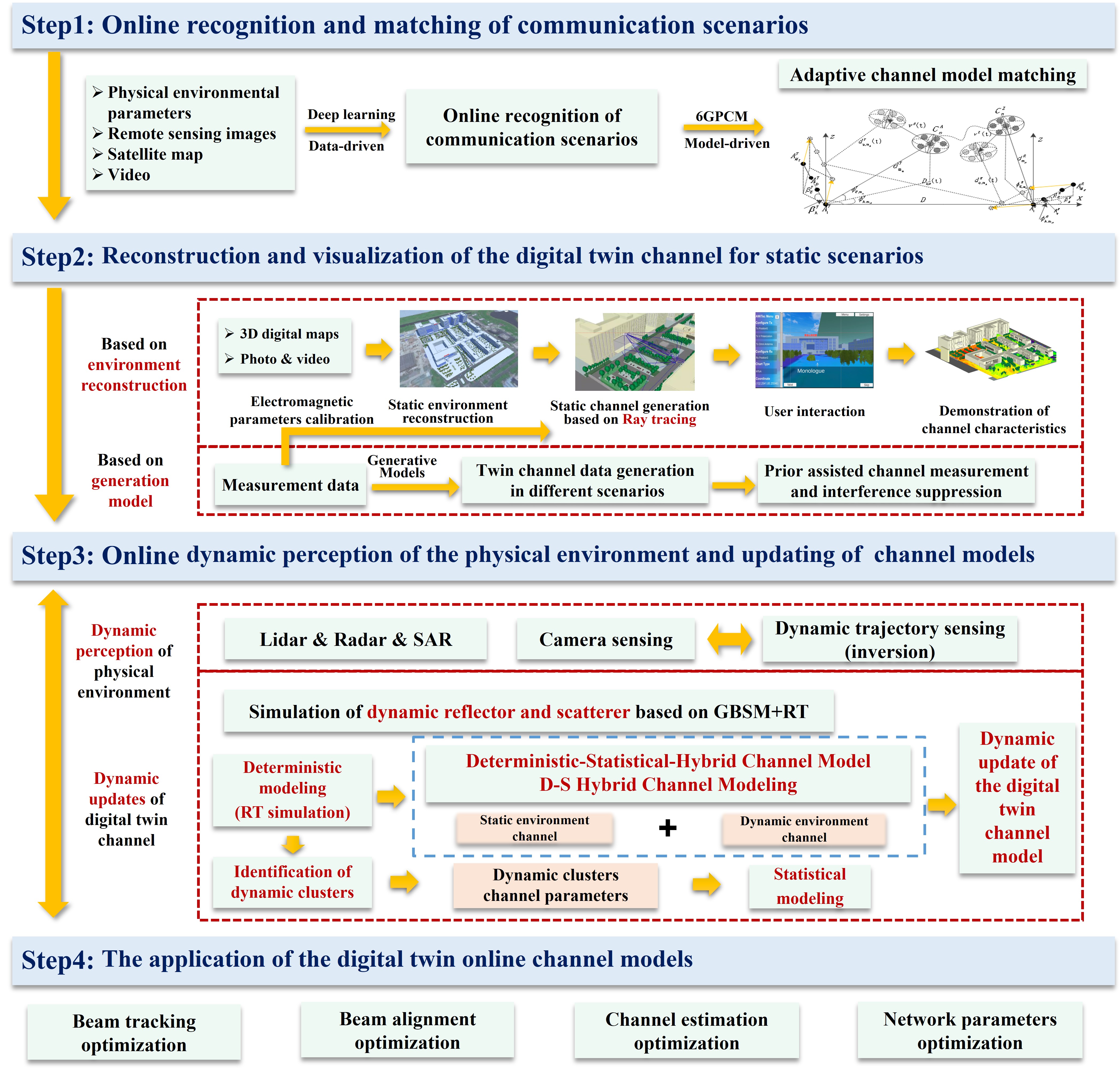}
    \caption{Main four steps for constructing the proposed DTOCM.}
    \label{steps}
\end{figure*}

\subsection{Machine Learning (ML)-Assisted Scenario Identification}
The current 5G standardized channel model has a relatively rough division of communication scenarios, which is insufficient to describe all communication scenarios under the vision of a full frequency band, full coverage, and full application of future 6G networks. To construct real-time and accurate DTOCM for all frequency bands and scenarios, it is necessary to classify 6G wireless communication scenarios in detail and explore the channel characteristics of different communication scenarios.

As shown in Fig. \ref{steps}, the first step in constructing DTOCM is to create comprehensive, detailed, and hierarchical classifications of 6G communication scenarios, followed by matching the appropriate channel model parameters. Specifically, based on the environmental perception data, a data-driven machine learning approach is used to identify and model communication scenarios as follows:
\begin{itemize}
    \item \textbf{Real-time environment perception}: To achieve comprehensive communication scenario recognition, various environmental data sources, such as electronic maps, remote sensing images, and point cloud data, are used to provide the real-time perception of the environment \cite{Fzhang}\cite{Koivumaki}. These sources facilitate the extraction of environmental parameters, which are critical for accurate channel modeling;
    \item \textbf{Parameter extraction and scenario classification}: The environmental parameters are then extracted and processed through machine learning models (such as neural networks), which classify communication scenarios into different categories, such as aerospace communication, airborne communication, terrestrial communication, and maritime communication. These categories are further broken down into more specific environments like satellites, space stations, aircraft, drones, indoor, urban, maritime, and islands. The neural network-based classification network is key to this process, allowing for a refined understanding of the various environments;
    \item \textbf{Channel model assignment}: Finally, different communication scenarios are identified and assigned with different channel model parameters of the 6GPCM \cite{6GPCM}. This involves mapping the appropriate channel model parameters to the scenarios, allowing for accurate multi-scenario channel measurements and ensuring compatibility with 6G channel models.
\end{itemize}


\subsection{Environment Reconstruction for Offline Channel Map Initialization}
\label{offline_map}
Based on the scenario identification result, the next step is reconstructing the digital twin environment for offline channel map initialization. This can be achieved by first using multimodal environmental perception, such as images, videos, 3D electronic maps, and point clouds, to conduct 3D scene reconstruction. The reconstructed virtual scenes are then imported into RT software to simulate the channel properties. By comparing the differences between the RT simulation results and the actual measurement data, the electromagnetic coefficients used in the simulation can be calibrated to improve the accuracy of the RT channel reconstruction. Apart from the twin environment, the digital twin layer also contains the twin data. Generative models can be used to create twins of different scenes from measurement data, providing a priori assistance for channel measurement and interference suppression.

\subsection{Environment Perception for Online Channel Map Update}
\label{dynamic_map}
The last step is to enable real-time updates of dynamic environmental channel characteristics. We use physical environment sensors, such as camera sensing, lidar, radar, and synthetic aperture radar (SAR) systems, to monitor environmental dynamic changes in real time, such as moving vehicles and pedestrians. Then, these changes will be reflected in the twin environment and channel characteristics will be re-simulated corresponding to such changes. The DTOCM employs a novel static-dynamic hybrid channel modeling algorithm to perform dynamic channel information updates. The static objects in the environment are simulated using RT modeling (as elucidated in Section \ref{offline_map}). For the dynamic objects, we use the parameters of dynamic clusters extracted from the measurement data and employ GBSM to simulate dynamic scattering channels.


\begin{figure*}[!t]
  \centering
  \includegraphics[width=0.9\textwidth]{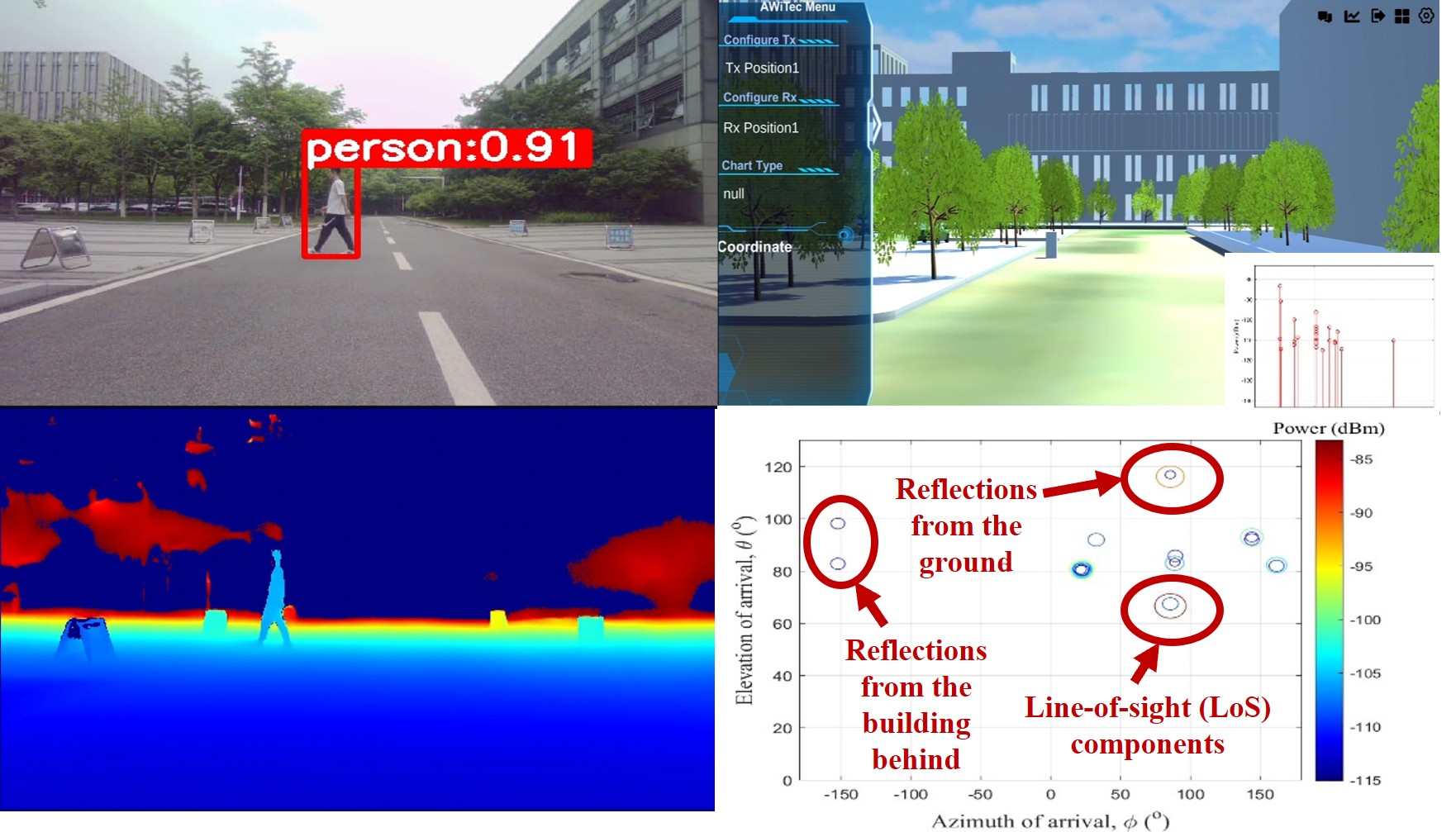} 
    \caption{Real-time channel information visualization of DTOCM.}
    \label{Visualization}
\end{figure*}
\label{Sec-visual}

\subsection{Applications of DTOCM}
The digital twin channels constructed based on the aforementioned steps can serve various application scenarios, enhancing specific aspects of 6G communication networks, such as beam alignment, beam tracking, channel estimation, and network parameters optimization \cite{beam2020}\cite{radio2021}.

\subsubsection{Beam alignment optimization} 
Beam alignment can be optimized by using channel information derived from DTOCM, which is established through the scenario identification and environment reconstruction in Steps 1 and 2. It supports the selection of the optimal performance beam, minimizing the effort and increasing the precision in aligning beams with the best signal quality. The DTOCM aids in pre-selecting these beams based on power spectral density and other relevant metrics, reducing the pilot overhead and complexity of channel estimation.

\subsubsection{Beam tracking optimization} 
DTOCM can be used to enhance beam tracking by generating channel statistics that optimize the detection beam's scope, thereby increasing both the speed and accuracy of beam tracking. By incorporating real-time environmental updates from Step 3, the DTOCM provides up-to-date channel information. Through employing channel sparsity via sparse Bayesian learning, beams are selected with the maximum expected signal-to-noise ratio, facilitating rapid beam alignment and efficient tracking. This optimization leads to faster response times in dynamic environments where quick adaptation to changes is crucial.

\subsubsection{Channel estimation optimization} 
The optimized channel estimation can use DTOCM to guide pilot design and frontload channel information acquisition steps, which significantly cuts down on the channel estimation overhead. By relying on the preliminary data from the comprehensive channel maps generated from Steps 1 to 3, the reduction of the complexity of channel estimations can be achieved and the overall efficiency of 6G communication networks is expected to be improved.

\subsubsection{Network parameters optimization} 
Unlike traditional channel modeling, DTOCM not only aims to characterize channel properties within specific scenarios, historical time data, and known frequency bands, but also to predict channel properties in unknown environments, future time, and unknown frequency bands. This predictive ability is enabled by the ML-assisted scenario identification and environment perception. This allows the prediction of future channel twin data based on perceived data from wireless propagation channels, which is then fed back to the network layer, allowing for network parameters optimization based on CSI.

\begin{figure}[h]
	\centering
	\subfigure[]{
        \includegraphics[width=0.45\textwidth]{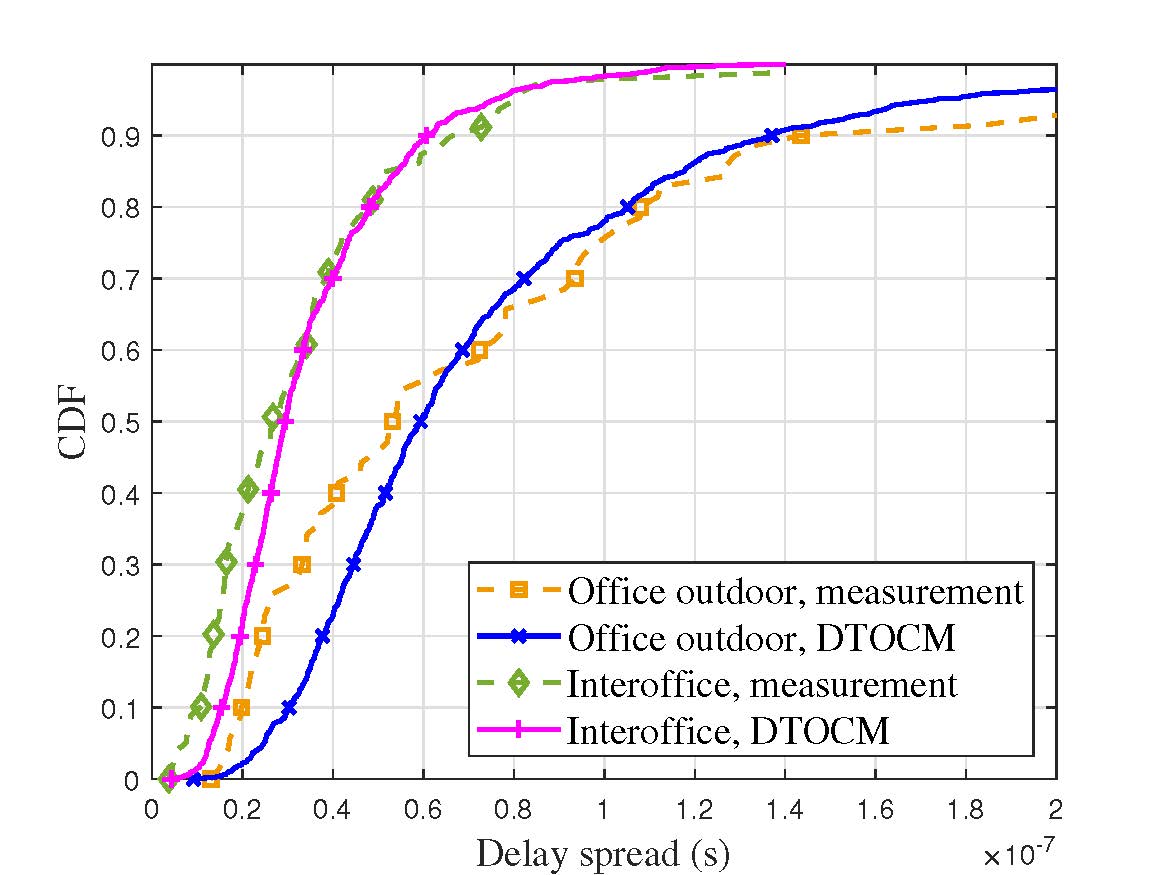}
	}
	\subfigure[]{
	\includegraphics[width=0.45\textwidth]{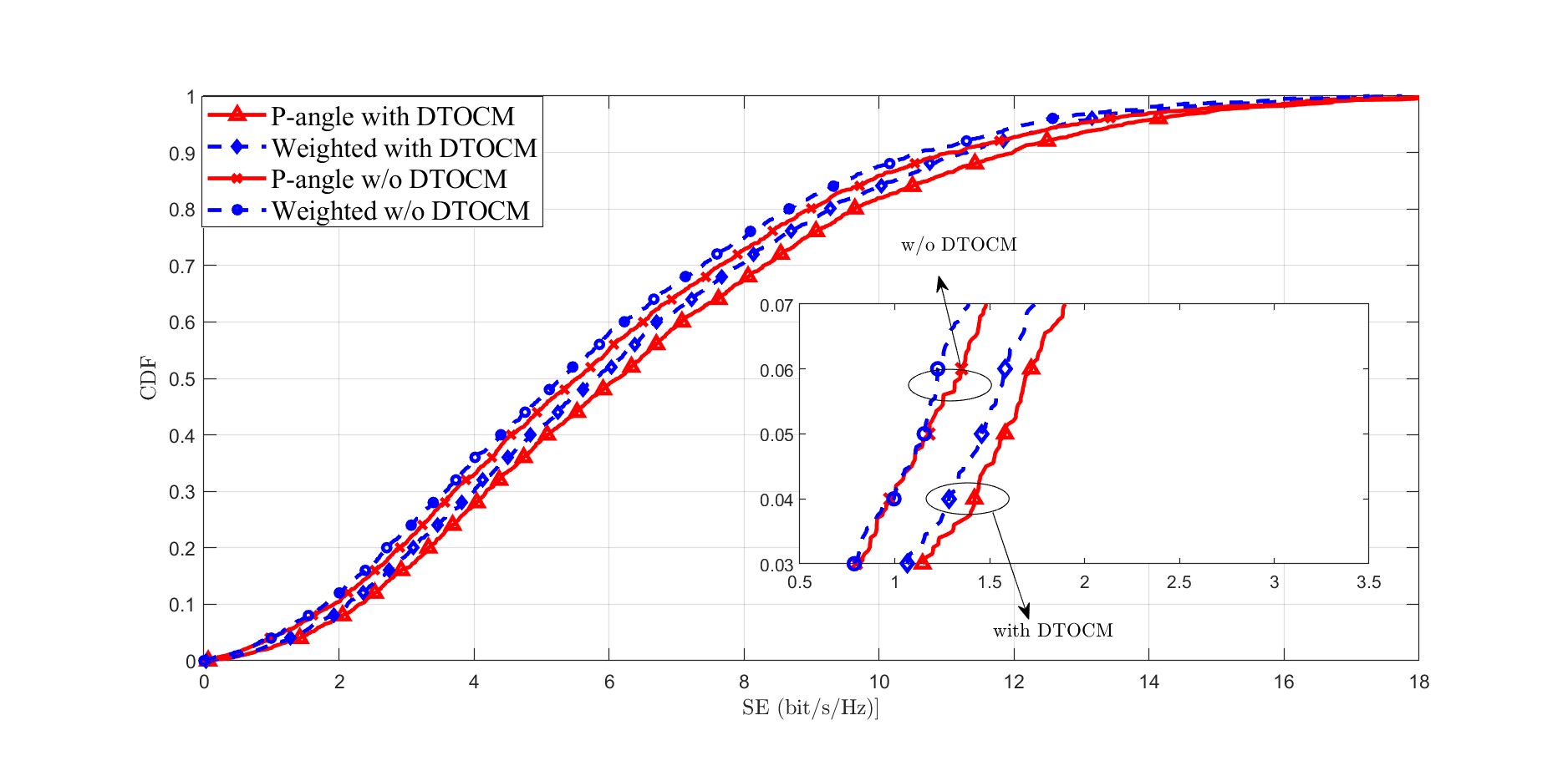}
	}
 \caption{Real-world performance demonstration of DTOCM: (a) Comparison of CDFs of DSs between DTOCM and measurements in two different scenarios. (b) CDFs of SE per user with and without DTOCM for different multiple access schemes.}
 \label{sim_results}
\end{figure}

\section{Real-Time Channel Information Provisioning and Visualization of DTOCM}
The proposed digital twin online channel modeling framework features real-time provisioning and visualization of channel information in 6G networks, which enables network performance analysis and intelligent network resource scheduling. 
The demonstration platform starts with the reconstruction of scenario environments. Static scenarios are built using Blender to model the physical features of the ``Wireless Valley", while dynamic scenarios involving moving objects like UAVs or persons are developed using Unity. This allows for interactive simulation where users can select and modify transmitter (Tx) and receiver (Rx) positions with mouse clicks, enhancing the realism and applicability of the model. The system outputs detailed channel characteristics, including CSI, path loss, and other relevant data. These outputs are dynamically updated based on user interactions and scenario changes. The modeling leverages RT and GBSM to accurately predict and simulate the wireless channel behavior in real time. To synchronize the acquisition of RGB and depth images, we utilize the Orbbec Gemini 2 XL camera, which offers a working distance of over 20 meters and provides high-quality RGB-depth images with precise synchronization capabilities. The computational platform consists of an Intel(R) Xeon(R) Platinum 8336C CPU @ 2.30GHz and an NVIDIA GeForce RTX 4090 GPU, enabling efficient processing of RGB and depth image data and supporting real-time computation and algorithm execution.

As illustrated in Fig. \ref{Visualization}, in the top left quadrant, real-time environmental changes are perceived, with a person detected with a confidence level of 0.91. The top right quadrant showcases the corresponding digital twin environment and the angular power spectral density, which updates dynamically to reflect the movements of the person in real time. The bottom left quadrant displays depth analysis through a depth map, indicating distances and depths with varying colors. The bottom right quadrant presents the corresponding channel characteristics, including azimuth and elevation angle PSDs, providing detailed insights into the wireless channel properties.

Fig. \ref{sim_results} shows two case studies demonstrating the real-world performance of DTOCM. Fig. \ref{sim_results}(a) compares the CDFs of delay spread (DS) between DTOCM and measurements in office outdoor and interoffice scenarios. It can seen that, for both scenarios, the simulation results from DTOCM have good agreements with the measurement results. This demonstrates the importance of ML-assisted scenario identification and the superior accuracy of DTOCM. In Fig. \ref{sim_results}(b), two multiple access schemes in angle domain and space domain are considered for comparison, namely ``P-angle" and ``Weighted", respectively. Fig. \ref{sim_results}(b) depicts the CDFs of the spectral efficiency (SE) per user. It can be observed that P-angle and Weighted with DTOCM both outperform over those without DTOCM.



\section{Open Research Issues}
\label{open_issue}
Despite the versatile applications of DTOCM in promoting 6G network management, there exist several critical challenges. In this section, we discuss some open research issues related to DTOCM for future 6G networks.

\subsection{High-fidelity RT Simulation}
The fidelity of RT simulations is a crucial component of DTOCM, which is affected by three main factors. \emph{First}, the accuracy of RT simulations is significantly influenced by the complexity and detail of the 3D models utilized. However, constructing highly detailed 3D models requires intensive computational resources and robust data inputs, posing a challenge in dynamic environments where changes occur frequently. \emph{Second}, the electromagnetic properties of the materials represented in the models also play a critical role in RT accuracy. Correctly assigning material characteristics such as reflectivity, absorbency, and permittivity is crucial, as these properties determine how waves interact with surfaces. \emph{Third}, the underlying algorithms used in RT simulations affect the fidelity of the results. Enhancements in the algorithms' ability to handle complex interactions like diffraction, scattering, and multiple reflections are necessary to increase simulation precision. The computational efficiency of these algorithms also impacts the feasibility of performing real-time simulations, which are essential for DTOCM.

\subsection{Perception and Fusion of Multimodal Data}
Another key challenge in advancing DTOCM is the effective perception and integration of multimodal data from diverse sources. This involves not only gathering static data from sensors and maps but also dynamically incorporating real-time data from mobile entities such as drones and vehicles. Integrating such varied data types—ranging from high-resolution images, real-time video feeds, and sensor data to 3D point clouds—requires sophisticated fusion algorithms capable of creating a cohesive and actionable channel model.

\subsection{Real-time Processing and Latency Reduction}
Addressing the challenges related to latency and real-time processing is also essential for ensuring the accuracy and responsiveness of DTOCM. Effective real-time integration of dynamic data from diverse sources such as sensors, drones, and user equipment is crucial. Additionally, minimizing latency in environmental perception and data processing is critical for the timely update of channel maps. This requires the development of low-latency processing algorithms and potentially adopting edge computing solutions that process data closer to the source. Addressing these challenges will enhance the DTOCM’s ability to provide accurate, real-time CSI which is crucial for dynamic 6G environments.


\subsection{Application to a 3D Continuous-Space Radio Environment}
For future networks, channels are evolving from discrete local space wireless propagation channels to 3D continuous-space radio channels. Applying DTOCM to a 3D continuous-space radio environment in 6G networks introduces several unique challenges. These include the complexity of the continuous 3D propagation environment, which encompasses spatiotemporal and frequency multi-scale channel electromagnetic characteristics. Additionally, the lack of geographic information and extensive channel measurement experiments, coupled with the scarcity of related data, makes it difficult to validate the accuracy of the models. Furthermore, the varying requirements for channel characterization across different transmission technologies introduce additional difficulties to applying DTOCM in 3D continuous-space radio environments.

\section{Conclusions}
\label{conclusion}
In this article, we have proposed a framework and design procedure for DTOCM, which can accurately characterize dynamic channels and greatly assist 6G network optimization. We have outlined the visions, associated challenges, construction procedures, and operational principles of DTOCM, emphasizing its capability for real-time channel information provisioning and visualization. A case study has been presented to highlight the superior accuracy of DTOCM in comparison with measurement data, addressing the mismatch issue between the synthetic channel in the lab simulation and the practical wireless channel in the real network. We hope that the findings in this article will spark further research and development in DTOCM, ultimately improving the efficiency, adaptability, and performance of future networks.

\section*{Author Information}

\begin{IEEEbiography}
[{\includegraphics[width=1in,height=1.25in,clip,keepaspectratio]{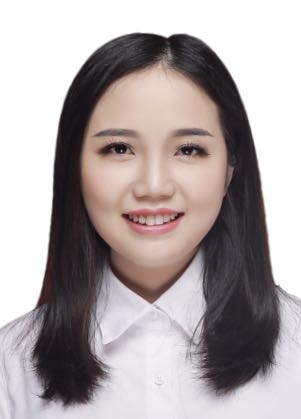}}]
{Junling Li} (junlingli@seu.edu.cn) received the B.S. degree from Tianjin University, Tianjin, China, and the M.S. degree from the Beijing University of Posts and Telecommunications, Beijing, China, in 2013 and 2016, respectively. In 2020, she received the Ph.D. degree from the University of Waterloo, Canada. She is currently an Associate Professor in the National Mobile Communications Research Laboratory at Southeast University, Nanjing, China. Her research interests include AI-based channel modeling and digital twin online channel modeling.
\end{IEEEbiography}

\begin{IEEEbiography}
[{\includegraphics[width=1in,height=1.25in,clip,keepaspectratio]{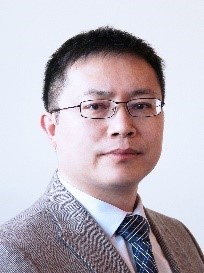}}]
{Cheng-Xiang Wang} (chxwang@seu.edu.cn) is a professor and the executive dean of the School of Information Science and Engineering, Southeast University, Nanjing, China. He is also a part-time professor with Purple Mountain Laboratories, China. His current research interests include wireless channel measurements and modeling, 6G wireless communication networks, and electromagnetic information theory. He is a Member of the Academia Europaea, a Fellow of the Royal Society of Edinburgh, and a Fellow of the IEEE.
\end{IEEEbiography}

\begin{IEEEbiography}
[{\includegraphics[width=1in,height=1.25in,clip,keepaspectratio]{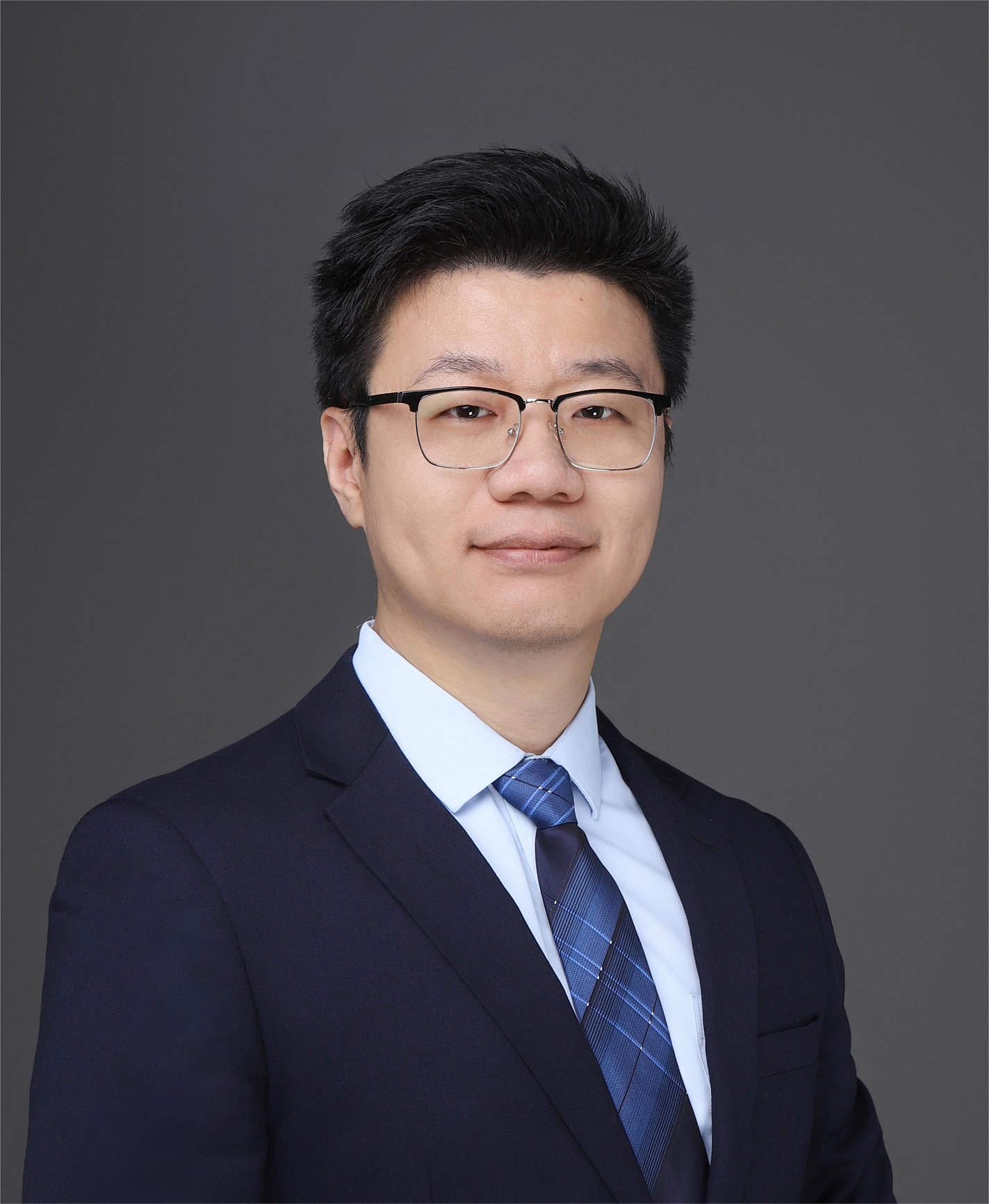}}]
{Chen Huang} (huangchen@pmlabs.com.cn) received the Ph.D. degree from Beijing Jiaotong University, Beijing, China, in 2021. From 2018 to 2020, he has been a Visiting Scholar with the University of Southern California and with the Universit\'e Catholique de Louvain. Since April 2023, he has been an associate professor in Purple Mountain Laboratories, and an extramural supervisor in Southeast University. His research interests include AI-based channel measurements and modeling.
\end{IEEEbiography}

\begin{IEEEbiography}
[{\includegraphics[width=1in,height=1.25in,clip,keepaspectratio]
{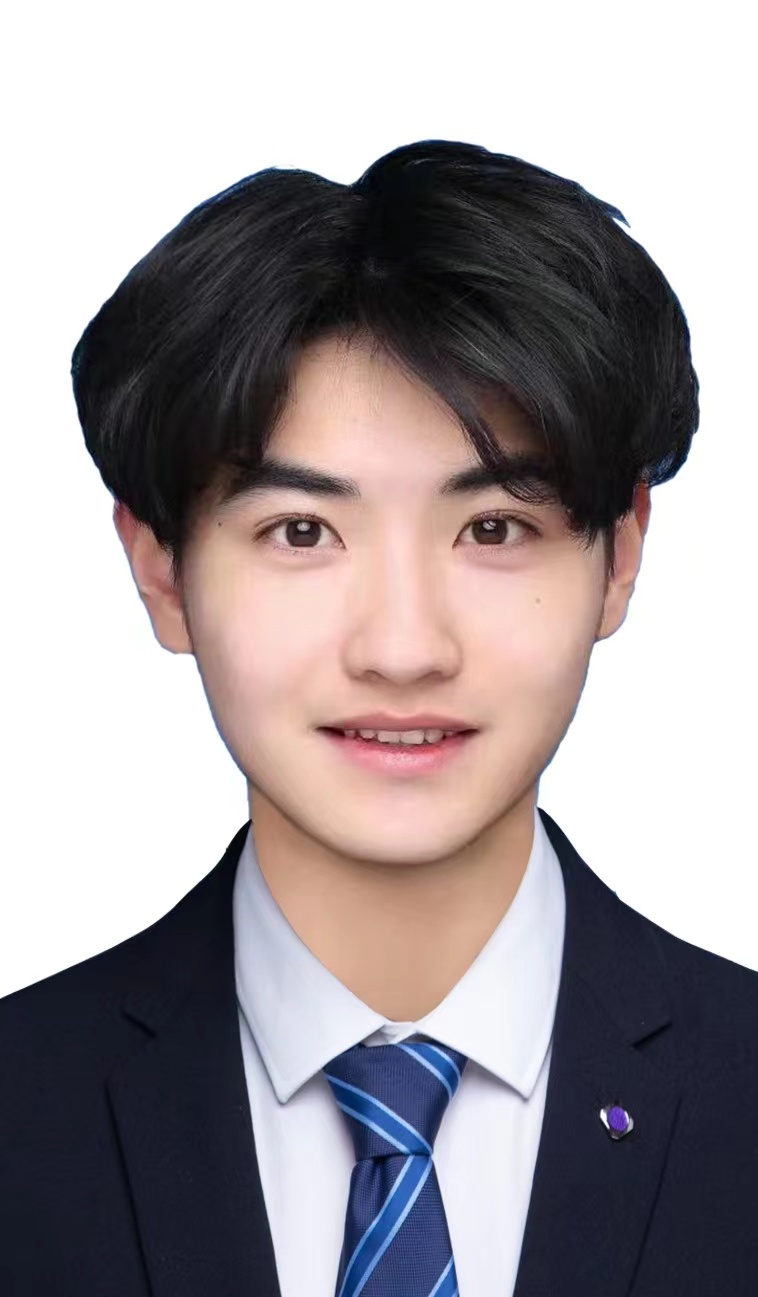}}]
{Tianrun Qi} (tr\_qi@seu.edu.cn) received the B.E. degree in Information Engineering from Southeast University, Nanjing, China, in 2022. He is currently pursuing the Ph.D. degree in the National Mobile Communications Research Laboratory, Southeast University, China. His research interests are 6G channel modeling and 6G dynamic channel map.
\end{IEEEbiography}

\begin{IEEEbiography}
[{\includegraphics[width=1in,height=1.25in,clip,keepaspectratio]
{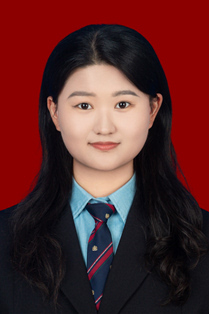}}]
{Tong Wu} (wutong@seu.edu.cn) received the B.S. degree from Southeast University, Nanjing, China, in 2023. She is currently pursuing the Ph.D. degree in Information and Communication Engineering with the School of Information Science and Engineering, Southeast University, Nanjing, China. Her research interest is predictive channel modeling based on image processing and machine learning.
\end{IEEEbiography}

\bibliographystyle{IEEEtran}

\end{document}